\documentclass[preprint,twocolumn,tight]{aastex701}

\usepackage{graphicx}
\usepackage{natbib}
\usepackage{amsmath}

\begin{document}

\title{VLBI Detections of Compact Nuclei in Spiral-hosted Double-lobed Radio-loud Active Galactic Nuclei (DRAGNs): Evidence for Weak Parsec-Scale Jet Activity}

\author[0009-0004-1075-5789]{Mingyu Ryu}
\affiliation{Department of Physics, College of Natural Sciences, Ulsan National Institute of Science and Technology (UNIST), 50 UNIST-gil, Eonyang-eup, Ulju-gun, 44919 Ulsan, Republic of Korea}
\email{}

\author[0000-0001-8229-7183]{Jae-Young Kim} 
\affiliation{Department of Physics, College of Natural Sciences, Ulsan National Institute of Science and Technology (UNIST), 50 UNIST-gil, Eonyang-eup, Ulju-gun, 44919 Ulsan, Republic of Korea}
\email[show]{jaeyoungkim@unist.ac.kr}

\begin{abstract}
We report milliarcsecond-scale VLBI detections of compact radio nuclei in four spiral-hosted, double-lobed radio-loud AGNs (spiral DRAGNs), a rare class that challenges the traditional association of powerful jets with elliptical hosts. Using public VLBI data archives, we identify compact cores in four sources and resolve parsec-scale jets in two of them. The VLBI components show low brightness temperatures ($T_{\rm b} \approx 10^9$\,K in the core) and jet-to-counterjet ratios consistent with only mildly relativistic intrinsic speeds ($\beta \lesssim 0.6$ for inclinations $\theta \lesssim 80^\circ$), indicating weakly powered pc-scale outflows. The low radio-Eddington ratios $\log(L_{\rm R,1.4\,GHz}/L_{\rm Edd}) \approx -5$ to $-8$ support this interpretation. Three objects lie on the fundamental plane of black hole activity, implying that global accretion-jet coupling in spiral DRAGNs is similar to that in other AGNs. Comparison with recent GRMHD simulations of thin-disk jets suggests that the VLBI-scale cores in spiral DRAGNs may trace an early or intermittently magnetized phase of jet launching. The coexistence of weak pc-scale jets and large kpc-scale lobes implies recurrent or long-duty-cycle jet activity in these late-type hosts.
\end{abstract}


\section{Introduction}\label{sec:intro}

Giant radio lobes of kpc-/Mpc-sizes have traditionally been associated with massive elliptical galaxies \citep{Urry95, Wilson95, McLure04, Best05}. 
These empirical findings have led to the idea that elliptical galaxies provide a natural environment to launch and collimate powerful jets from their central supermassive black holes (SMBHs) (see \citealt{Blandford19} for a detailed review). 
In more detail, elliptical galaxies often host hot and pressure-supported accretion flows in which the gas forms a geometrically thick disk. 
These flows, commonly referred to as advection-dominated accretion flows (ADAFs) \citep{Ichimaru77,Narayan94,Narayan95}, are considered as a key ingredient for the generation of vertical magnetic flux through the ergosphere of central spinning SMBHs and external pressure support to collimate and magnetically accelerate the jet up to galactic scales \citep{Komissarov07,Kim18,Blandford19}. 
This framework has provided a natural explanation for the observed dichotomy about spiral galaxies tend to host weak jets because of their different host environments, compared to the ellipticals.

However, there has been an increasing number of discoveries of rare spiral and disk galaxies that host extended kpc-scale double-lobed radio structures. Being referred to as spiral double-lobed radio-loud AGNs (spiral DRAGNs), these objects challenge the aforementioned framework of jet formation and collimation by the environmental effects \citep{Ledlow98,hota11,mao15,Singh15,vietri22,Wu22}. 
For instance, \cite{Ledlow98} provided one of the first pieces of firm evidence for a powerful double-lobed radio AGN in a serendipitous spiral host galaxy 0313$–$192.  
Subsequently, \cite{hota11} reported another spiral-hosted radio galaxy exhibiting multiple episodic jet activities in 345\,MHz images from the Giant Metrewave Radio Telescope (GMRT).
\cite{Singh15} further reported four spiral galaxies with kpc-scale radio lobes, all of which were classified as Fanaroff-Riley type II (FR II) radio galaxies whose lobes extended to over 80\,kpc-scales despite their moderate radio luminosities $(\sim 10^{24}-10^{25} \rm \,W \,Hz^{-1})$ at 1.4\,GHz. 
More recently, \cite{Wu22} significantly expanded the total number of known spiral DRAGN candidates, using high-resolution Hubble Space Telescope (HST) optical and Very Large Array (VLA) radio images, focusing on a sample derived from the Gem of Galaxy Zoo project \citep{Keel22}.

These discoveries led to several hypotheses about how spiral galaxies might generate powerful jets similar to those in elliptical galaxies. 
On one hand, \cite{Ledlow98} suggested that strong galaxy–galaxy interactions, such as major mergers, can trigger the AGN activity, and \cite{Mao18} suggested that a lower-density interstellar medium may help the jets expand more easily into kiloparsec-scale halos.
On the other hand, \cite{Wu22} argued that the likelihood and strength of radio jets depend primarily on the stellar mass of the bulge and thus the mass of the central SMBH, rather than on galaxy morphology. 
Unfortunately, only a limited amount of observational studies have been made in this regard, making it difficult to identify which of these scenarios would be more realistic to understand the origin of spiral DRAGNs.

In this regard, VLBI detection and imaging of compact radio nuclei of spiral DRAGNs offer two possible key improvements. 
First, the relatively poor angular resolutions (e.g., of order of arcseconds) of the optical and radio images used in previous studies still suffer from systematic errors in the optical-radio image alignment, affecting the statistical significance of the optical-radio association 
(see \citealt{Wu22}).
In contrast, VLBI-scale high-resolution radio images naturally result in significantly improved source position accuracy, and therefore can provide bona-fide optical-radio associations.
Second, high-resolution images of the radio nuclei and (if present) extended radio jets can allow measurements of the geometry of the outflow structure (e.g., \citealt{Lee25}) and jet plasma properties close to the central engine, using observables such as the brightness temperature and intrinsic jet speed. 
For example, blazar-like jets typically exhibit fast bulk motions, with typical bulk Lorentz factors of the order of $\Gamma \gtrsim 10$, as inferred from the population modeling of a number of blazar and quasar jets \citep{Lister19}.
In contrast, observational constraints the ranges of $\Gamma$ for spiral DRAGNs are lacking. 
Also, along with the collimation profile \citep{Lee25}, the dynamical properties and ultimately the total power of jets in spiral DRAGNs can be compared with jets in more diverse sources such as blazars, quasars, and radio galaxies, which can provide hints about the jet formation in those serendipitous objects. 

To this end, we aim to expand the number of VLBI-detected spiral DRAGNs and analyze physical properties of the compact radio nuclei and jets, if present. 
Accordingly, this paper is structured as follows.
In Sect.~\ref{sec:data}, we describe our targets and datasets used for our study, including a high-probability spiral DRAGN sample and archival VLBI data from various databases. 
In Sect.~\ref{sec:result}, we present our main results about a list of spiral DRAGN objects with reliable VLBI detections, morphologies of the compact radio nuclei and jets, measurements of the jet brightness temperatures, calculations of the radio-Eddington ratios, and constraints on the intrinsic jet speeds. 
We also compare the black hole masses, X-ray luminosities, and VLBI-scale radio luminosities with those of well-established jet-powered AGNs, in order to see if spiral DRAGNs lie well on the so-called fundamental plane of active  black holes \citep{Merloni03}.
In Sect.~\ref{sec:discussion}, we finally compare these results with those of well-known powerful jets in quasars and blazars, and discuss the possible origin and dynamics of jets in the spiral DRAGN objects. 
Throughout the paper, we adopt a $\Lambda$CDM cosmology with 
$H_{0} = 67.8\,\mathrm{km\,s^{-1}\,Mpc^{-1}}$, 
$\Omega_{m} = 0.31$, and 
$\Omega_{\Lambda} = 0.69$ 
\citep{PlanckCollaboration14}.

\section{Data and Analysis}\label{sec:data}

\begin{sidewaystable*}[t!]
\centering
\setlength{\tabcolsep}{3pt}
\caption{Properties of the radio images of the targets.} 
  \begin{tabular}{ccccccccccc}
    \hline \hline
    Target & R.A. & Dec. & Array & Obs. date & $\nu$ & Beam size & Beam P.A. & Peak intensity & Total flux density & $\sigma$\\
    (J2000) & ($^h\ ^m\ ^s$) & (° $'$ $''$) & \empty & (yyyy/mm/dd) & (GHz) & ($'' \times ''$) & (deg) & (mJy/beam) & (mJy) & (mJy/beam)\\
    \hline
    J0219+0155 & 02 19 58.7 & +01 55 49 & VLA & 2009/03/29 & 1.5 & 6.4$\times$5.4 & 0.00 &   62$\pm$6 & 317$\pm$32 & 0.137 \\
    \empty & \empty & \empty & VLBA & 2010/09/14 & 8.6 & 0.0047$\times$0.0011 & -14.8 &   22$\pm$2 & 21$\pm$2 & 0.660 \\
    \hline
    J1159+5820 & 11 59 05.8 & +58 20 36& LOFAR & 2016/04/04 & 0.144 & 6.0$\times$6.0 & 0 & 1.7$\pm$0.2 & 1773$\pm$177 & 0.093 \\
    \empty & \empty & \empty & VLBA & 2006/05/27 & 4.8 & 0.0028$\times$0.0021 & -0.28 & 62$\pm$6 & 297$\pm$30 & 0.300 \\ 
    \hline
    J1352+3126 & 13 52 17.8 & +31 26 46 & LOFAR & 2017/02/09 & 0.144 & 6.0$\times$6.0 & 0 & 8636$\pm$864 & 20691$\pm$2069 & 1.5 \\
    \empty & \empty & \empty & VLBA & 2010/09/11 & 8.6 & 0.0025$\times$0.0011 & 19.9 & 34$\pm$3 & 69$\pm$7 & 0.220 \\
    \hline
    J1649+2635 & 16 49 23.9 & +26 35 03 & VLA & 1995/11/13 & 1.4 & 5.4$\times$5.4 & 0.00 & 6.2$\pm$0.6 & 112$\pm$11 & 0.757 \\  
    \empty & \empty & \empty & VLBA & 2012/02/29 & 8.4 & 0.0037$\times$0.0009 & 28.0 & 16$\pm$2 & 17$\pm$2 & 0.400 \\
    \hline \hline
  \end{tabular}
  \tablecomments{
  From left, each column shows 
  the source names in J2000, 
  source coordinates in R.A. and Dec., 
  observing radio arrays,
  observing dates in yyyy/mm/dd format,
  observing frequencies in GHz,
  beam sizes in arcsecond,
  position angle of the major axis of the beam in degrees from north to east,
  peak intensity in mJy/beam,
  total flux density in mJy,
  and the image rms noises in mJy/beam.
  Here, systematic flux uncertainties of 10\% are adopted for the peak intensities and total flux densities.
  } 
  \label{tab:VLBA_data_summary}
\end{sidewaystable*}

\begin{table*}[t!] 
    \caption{Optical images retrieved from SDSS and HST archives.}
    \centering
    \begin{tabular}{cccc}
        \hline
        Source      & Database & Band          & ObsID or ObjID                               \\ \hline
        J0219+0155 & SDSS DR 18     & r             &
        1237678618506887332                  \\ \hline
        J1159+5820  & SDSS DR18      & r             & 1237661354316726385            \\ \hline
        J1352+3126  & SDSS DR18      & r             & 1237665331471515679            \\ \hline
        J1649+2635  & SDSS DR18      & r             & 1237662301913940522            \\ \hline
        J0219+0155  & HST (ACS/WFC)  & F475W         & hst\_15445\_3y\_acs\_wfc\_f475w\_jds43y       \\ \hline
        J1352+3126  & HST (WFPC2)    & F702W         & u27l5a02t                             \\ \hline
    \end{tabular}
    \tablecomments{
  From left, each column shows the source names in J2000, the databases from which the image was retrieved, the filter bands, and the corresponding observation or object IDs from the HST or SDSS databases, respectively.}
  \label{tab:observation_IDs_of_optical_images}
\end{table*}

\subsection{VLBI spiral DRAGN Sample}\label{subsec:sample}

For the purpose of our study, we selected spiral DRAGNs with low probabilities $P$ of chance alignment between the optical galaxy and the radio lobes ($P\lesssim10^{-5}$), based on the catalog published by \cite{Wu22}. We also supplemented this list with additional individual spiral DRAGN objects reported in the other literature, including 
0313$-$192 \citep{Ledlow98}, 
J0354$-$1340 \citep{vietri22}, 
J0836+0532, J1159+5820, J1352+3126 \citep{Singh15}, J1030+5516 \citep{rakshit18}, 
J1409$-$0302 \citep{hota11}, 
J1649+2635 \citep{mao15}, and 
J2345$-$0449 \citep{bagchi14}. 
We refer to those literature for detailed discussions of their discovery and identification as spiral DRAGNs.
We then searched various public VLBI databases, including 
the Astrogeo VLBI FITS image database\footnote{\url{https://astrogeo.smce.nasa.gov/vlbi_images/}},
the European VLBI Network (EVN) data archive\footnote{\url{https://archive.jive.nl/scripts/portal.php}}, and 
the National Radio Astronomy Observatory (NRAO) data archive\footnote{\url{https://data.nrao.edu/portal/}}, in order to find any of the above-mentioned spiral DRAGNs that could have already been observed and detected by previous VLBI experiments.
Based on our search, we were able to find, from the Astrogeo database, four new objects with clear VLBI detections: J0219+0155, J1159+5820, J1352+3126, and J1649+2635
(excluding 0313$-$192 whose VLBI detection was already reported by \citealt{Mao18}). 
The data of J0219+0155, J1352+3126, and J1649+2635 were obtained as part of the VLBI 2MASS radio astronomy project \citep{condon11} using the Very Long Baseline Array (VLBA) (at 8.6\,GHz for J0219+0155, J1352+3126 and 8.4\,GHz for J1649+2635). Their fully calibrated visibility data were available from the Astrogeo database. 
We refer to \cite{condon11} for the acquisition, calibration, and imaging of the datasets.
The data of J1159+5820 were obtained as part of the VLBA Imaging and Polarimetry Survey (VIPS) program \citep{helmboldt07} using the VLBA at 4.8\,GHz. Also, fully calibrated data of this dataset were available from the Astrogeo database.
The corresponding source information and observing parameters are listed in Table \ref{tab:VLBA_data_summary}.
We also provide detailed descriptions of the individual sources in Sect.~\ref{sec:details_all_targets} of Appendix.
We note that a few other spiral DRAGN objects were also observed in VLBI by various instruments, but typically only as calibrators with limited numbers of scans (mostly single scan only). These data were excluded because of their limited $(u,v)$-coverages that could make it difficult to image and analyze the source structure and properties.

\begin{table*}[t]
  \setlength{\tabcolsep}{5pt}
  \centering
  \caption{
  Various physical properties of the four spiral DRAGNs. 
  }
  \begin{tabular}{ccccccccc}
    \hline \hline
    Target     & Redshift & $d_{L}$                 & $\log M_{\rm BH}$    & $\log L_{\rm X}$     & $\log L_{\rm R, 5\,GHz}$& $\log L_{\rm R, 1.4\,GHz}$  & $\log L_{\rm Edd}$ & $\log(L_{\rm R, 1.4\,GHz}/L_{\rm Edd})$ \\
    (J2000)    & (z)      & (Mpc)               & ($M_\odot$)    & (erg/s)        & (erg/s)         & (erg/s)            &  (erg/s) & \\
    \hline
    J0219+0155 & 0.041    & $187.0$    & 8.85$\pm$1.09  & $\*$ N/A       & $39.7\pm0.1$ & $39.1\pm0.2$ & $47.0\pm1.1$ & $-7.9\pm1.1$ \\
    \hline
    J1159+5820 & 0.054    & $248.6$    & 8.68$\pm$0.24  & 41.7$\pm$0.1 & $41.01\pm0.04$ & $41.29\pm0.04$ & $46.8\pm0.3$ & $-5.49\pm0.25$ \\
    \hline
    J1352+3126 & 0.045    & $205.8$    & 8.28$\pm$0.25 & 42.81$\pm$0.03 & $40.4\pm0.1$ & $40.24\pm0.04$ & $46.4\pm0.2$ & $-6.14\pm0.31$ \\
    \hline
    J1649+2635 & 0.055    & $253.4$    & 8.46$\pm$0.26 & $\leq$43.3 & $39.8\pm0.1$ & $39.3\pm0.2$ & $46.6\pm0.2$ & $-7.31\pm0.30$ \\
    \hline \hline
  \end{tabular}
  \tablecomments{
  From left, columns show 
  the target names in J2000, 
  the redshifts (\citealt{Wu22} for J0219+0155; \citealt{Singh15} for J1159+5820, J1352+3126, and J1649+2635), 
  luminosity distances computed from the redshifts in Mpc, 
  black hole masses in $M_{\odot}$, 
  $2-10$\,keV X-ray luminosities, 
  5\,GHz and 1.4\,GHz radio luminosities ($L_{\rm R,\nu} \equiv \nu L_\nu$), 
  Eddington luminosities, and 
  radio Eddington ratios.
  }
  \label{tab:properties_of_targets}
\end{table*}

\subsection{Kpc-scale radio images}\label{subsec:other_radio_data}

In order to check if deeper and higher resolution kpc-scale images of radio jets are available for the four targets from recent radio continuum surveys, 
we searched various surveys including 
the Faint Images of the Radio Sky at Twenty Centimeters (FIRST) survey at 1.4\,GHz with the Very Large Array (VLA) \citep{Becker95}, 
the Low Frequency Array (LOFAR) Two-Metre Sky Survey Data Release 2 at 144\,MHz (LoTSS DR2; \citealt{Shimwell22}), and 
the VLA Sky Survey (VLASS) at 3\,GHz \citep{Lacy20}.
The FIRST database provided images for all of our targets. LoTSS DR2 images were available for two of them (J1159+5820 and J1352+3126).
None of our sources were available from the VLASS database.
For the available images, we used the CARTA software \citep{carta} to measure the basic properties of the radio images such as the total flux densities.

\subsection{Optical images}\label{subsec:optical_images}

We also searched public archives for optical images of our four targets, primarily making use of the Canadian Initiative for Radio Astronomy Data Analysis (CIRADA\footnote{http://cutouts.cirada.ca/}) and Mikulski Archive for Space Telescopes (MAST\footnote{https://archive.stsci.edu/}) databases.
We found four Sloan Digital Sky Survey (SDSS) Data Release 18 \citep{Kollmeier26}
$r-$band images of all the four sources. 
In addition, we also found Hubble Space Telescope (HST) images of two of our targets from the MAST and Canadian Astronomy Data Centre (CADC\footnote{https://www.cadc-ccda.hia-iha.nrc-cnrc.gc.ca/en/} database: J0219+0155 and J1352+3126. 
The corresponding observation or object IDs are listed in Table \ref{tab:observation_IDs_of_optical_images}.
For J0219+0155 and J1352+3126, we adopted the available HST images to better display the host-galaxy structure, benefiting from their higher spatial resolution compared to the SDSS data.

\subsection{VLBI data analysis}\label{subsec:vlbi_properties}

We used the Difmap software \citep{Shepherd97} to image the sources from the visibility data, by using the CLEAN task in a standard manner for typical VLBI observations
(e.g., \citealt{Kim23}). 
Since all these data were already fully self-calibrated in phase and amplitude, we did not make additional self-calibrations during the imaging. 
After the imaging, we exported the CLEAN images outside the Difmap software and used the CARTA software to measure the peak intensities, total flux densities, and rms noise levels of the maps.

\begin{figure*}[ht!] 
    \centering
    \includegraphics[width=0.95\textwidth]{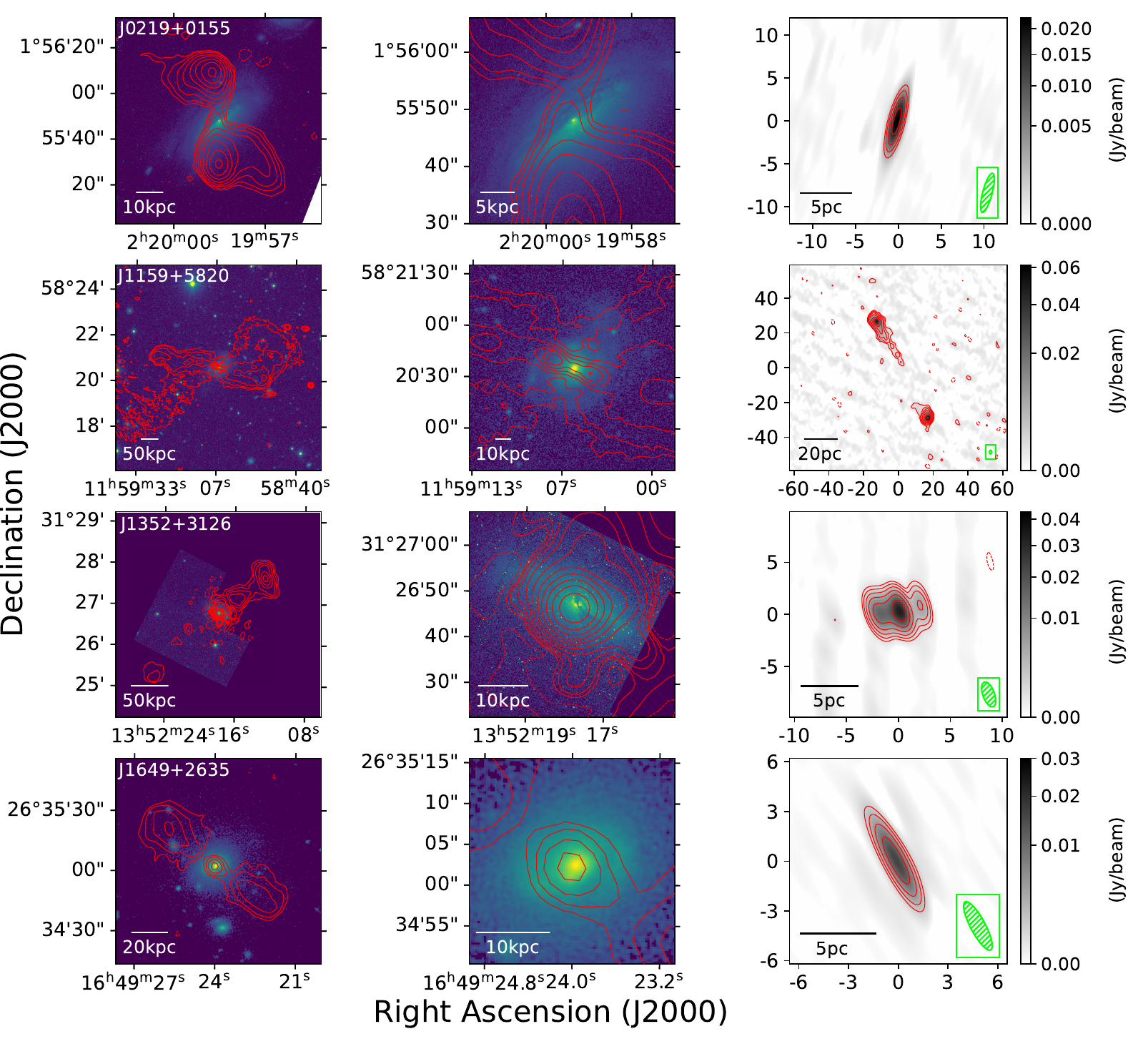}
    \caption{
    Optical and radio images of the four targets at various spatial scales. 
    HST (SDSS) optical images are shown for J0219+0155 and J1352+3126 (J1159+5820 and J1649+2635). 
    As for radio, LOFAR (VLA) images are shown for J1159+5820 and J1352+3126 (J0219+0155 and J1649+2635). 
    All the VLBI images are obtained by the VLBA.
    The left and center columns show kpc-scale radio (red contours) and optical (color, in $\sqrt{\rm intensity}$) images of the targets in absolute RA and Dec.
    The right column shows the parsec-scale VLBI images, showing the total intensities in both contours and color in relative RA and Dec. 
    The corresponding VLBI beam sizes are shown as green ellipses in the bottom right corner of each panel. 
    In each row, the source names are shown in the top left corner of the leftmost image.
    In all panels but kpc-scale radio structures of J1159+5820 and J1352+3126, contours are displayed at $(-1, 1, 2, 4, 8, 16, 32, 64, 128, 256) \times 3\sigma$ levels.
    For kpc-scale radio images of J1159+5820 and J1352+3126, higher starting levels of $4\sigma$ and $6\sigma$ are adopted, respectively.
    Bars at the bottom left corners of each panel show the corresponding spatial scales.
    }
    \label{fig:all_images}
\end{figure*} 

In order to parametrize physical properties of the VLBI-scale radio-emitting structures, 
we computed the observed brightness temperatures, $T_{\rm b}$, of the distinct emitting jet features. 

For this, we modeled the VLBI visibilities with circular Gaussian components using the Modelfit task in Difmap, in a similar manner as described in \cite{Kim25}.
For each component, we computed $T_{\rm b}$ as
\begin{equation}
T_{\rm b} = 1.22\times10^{12}\frac{F_{\nu}}{\nu^2 \Theta^2}~\rm K, 
\label{eq:tb}
\end{equation}
(e.g., \citealt{Kim18}) where 
$F_{\nu}$ is the flux density in Jy at an observing frequency,
$\nu$ in GHz,
and $\Theta$ is the Gaussian FWHM in mas.
We note that the intrinsic brightness temperature, $T_{\rm b,int}$, is related to $T_{\rm b}$ by $T_{\rm b,int}=T_{\rm b}\times (1+z)/\delta$ where 
$z$ is the redshift of the target,
and $\delta$ is the Doppler factor.
As for the model fitting procedure, we successively added increasing number of Gaussian components for each VLBI dataset, until iterative modelfit task reduced the image noise level comparable to the rms of the CLEAN maps and reasonable $\chi^{2}$ values were obtained (see also \citealt{kim19}).
After convergence, we computed the statistical uncertainties of the fitted parameters of each component as follows: 
\begin{equation}\label{eq:err}
\begin{aligned} 
\sigma_{\rm peak}=\sigma_{\rm rms}\left(1+\frac{I_{\rm peak}}{\sigma_{\rm rms}}\right)^{1/2},\\
\sigma_{\rm tot}=\sigma_{\rm peak}\left(1+\frac{F^{2}_{\rm tot}}{\sigma^{2}_{\rm peak}}\right)^{1/2},\\
\sigma_{d}=d\frac{\sigma_{\rm peak}}{I_{\rm peak}},\\
\sigma_{r}=\frac{1}{2}\sigma_{d},\\
\sigma_{PA}=\arctan\left(\frac{\sigma_{r}}{r}\right),
\end{aligned}
\end{equation}
\citep{fomalont99,lee08,schinzel12} where $\sigma_{\rm rms}$ is the image rms noise, $\sigma_{\rm peak}$, $\sigma_{\rm tot}$, $\sigma_{d}$, $\sigma_{r}$, $\sigma_{PA}$ are the uncertainties of the peak intensity $I_{\rm peak}$ (in mJy/beam), total flux density $F_{\rm tot}$ (in mJy), component FWHM size $d$ (in mas), radial distance to the component $r$ (in mas), and component position angle $PA$ (in radian), respectively.
Also, a systematic 10\% uncertainty was assumed for the total flux density, while for the FWHM size we assumed a 5\% uncertainty \citep{Kim25}.
Those systematic errors were added in quadrature to $\sigma_{\rm tot}$ and $\sigma_{d}$ and then propagated to estimate the total uncertainty of $T_{\rm b}$.

\subsection{BH masses, X-ray luminosities, radio luminosities, and radio-Eddington ratios}\label{subsec:other_mwl_data}

We also gathered additional information about the black hole masses, $M_{\rm BH}$, $2-10\rm~keV$ X-ray luminosities, $L_{\rm 2-10\,keV}$, and 5\,GHz radio luminosities, $L_{\rm R,5\,GHz}$, of the total VLBI-scale structures for our targets, in order to investigate the correlation between the three parameters and understand if spiral DRAGNs would follow the so-called fundamental plane of active jetted black holes \citep{Merloni03}. 
A summary of these values is given in Table~\ref{tab:properties_of_targets}.
For $M_{\rm BH}$, we adopted values for J1159+5820, J1352+3126, and J1649+2635 from \cite{Singh15}, where the black hole masses were estimated using the black hole mass–bulge luminosity relation \citep{Kormendy95,Magorrian98} for late-type galaxies \citep{McConnell13}.
For J0219+0155, we made use of the $M_{\rm BH}-M_{\rm *}$ relation in \cite{Wu22}, explicitly as follows: 
\begin{equation}
\log\left({M_{\mathrm{BH}}} \over {M_{\odot}}\right) = (7.43 \pm 0.09) + (1.61 \pm 0.12)\, \log\left({M_*} \over {M_0}\right),
\label{eq:m-sigma_relation}
\end{equation}
where $M_{\rm BH}$ is the mass of the black hole, $M_{0}=3\times10^{10}M_{\odot}$, and $M_{*}$ is the stellar mass of the host galaxy. 
As noted by \citet{Kormendy13}, the correlation between bulge properties and black hole masses could become highly uncertain for pseudo-bulge systems.
For this reason, we use the $M_{\rm *}$ for the host galaxy and, following \citet{Wu22}, adopt an intrinsic scatter of 0.81\,dex in the estimated $M_{\rm BH}$.
For $L_{\rm 2-10\,keV}$, we referred to various literature for the objects J1159+5820 \citep{Balasubramaniam20} and J1352+3126 \citep{Ueda05}. 
For J1649+2635, only an upper limit was available for $L_{\rm 2-10\,keV}$ \citep{Bohringer01}. 
For J0219+0155, no value of $L_{\rm 2-10\,keV}$ was reported in the literature.

The radio luminosity, $L_{\rm R,\,\nu} \equiv \nu L_{\nu}$, was derived from the total VLBI-scale flux density $F_{\nu}$ using $L_{\nu}=4\pi d_{L}^{2}F_{\nu}$ where $d_{L}$ is the luminosity distance with uniform systematic error of 10\% for measured total flux density. 
We estimated $F_{\rm 5\,GHz}$ using the 4.8 and 8.6\,GHz flux densities, assuming flat or steep spectra depending on the source morphologies--flat for core-dominated sources J0219+0155 and J1649+2635 and steep for jet-dominated sources J1159+5820 and J1352+3126, respectively.
For the core-dominated sources, we simply presume $F_{\nu} = F_{\rm 4.8\,GHz}=F_{\rm 8.6\,GHz}$.
Since the exact spectral indices of the sources are unknown, possible systematic uncertainties of $F_{\rm 5\,GHz}$ were estimated by assuming a power-law 
$F_{\nu} = F_{\nu_{0}} \left({\nu} / {\nu_{0}}\right)^{+\alpha},$
where $F_{\nu_{0}}$ is the flux density at an observing frequency $\nu_{0}$ and $\alpha$ is the spectral index. 
Then, the spread of $F_{\rm 5\,GHz}$ was computed for a range of spectral indices $\alpha \in (-1.0, 0.0)$. 
At $\nu_{0} = 8.6$\,GHz, the spread of $F_{\rm 5\,GHz}$ yielded a maximum possible difference of $\sim 21\%$ for $F_{\rm 5\,GHz}$, which we adopted as the systematic uncertainty. 
At $\nu_{0} = 4.8$\,GHz, the maximum difference was only $\sim 1\%$ and therefore was neglected. 
In both cases, we included an additional 10\% systematic uncertainty in the flux density, reflecting a possible data calibration error.
For the jet-dominated sources, we estimate $F_{\rm 5\,GHz}$ by applying a steep power-law spectrum and extrapolating the observed flux densities at 4.8 and 8.6\,GHz to 5\,GHz.
For J1159+5820, we adopt a steep spectral index of $\alpha = -1.5$, which is estimated by \cite{Tremblay16}, while for J1352+3126 we use a typical value of $\alpha = -0.7$. For J1352+3126, again a similar 21\% of systematic flux density error was adopted and added in quadrature.

To quantify the relative strength of the radio emission of the pc-scale VLBI core versus the Eddington luminosity, we calculated a monochromatic radio-Eddington ratio, $L_{\rm{R,1.4\,GHz}}/L_{\rm Edd}$. 
We adopted $\nu = 1.4\,\rm GHz$ as the reference frequency and extrapolated from observing frequencies of 4.8 and 8.6\,GHz to 1.4\,GHz, in order to enable direct comparisons of the ratios with values from recent studies (e.g., \citealt{King17, Ayubinia23}).
The $F_{\rm 1.4\,GHz}$ were derived from the total flux density measurements, following the same source-dependent assumptions about $\alpha$.  
As for the errors due to $\alpha$, now the possible uncertainties in $F_{\rm 1.4\,GHz}$ increases to a maximum possible difference in $F_{\rm 1.4\,GHz}$ of $\sim 42\%$ for the core-dominated sources. 
Similar steps were followed to estimate $F_{\rm1.4\,GHz}$ for the jet-dominated sources and the 42\% systematic uncertainty was assumed J1352+3126. 

\section{Results}\label{sec:result}

\begin{figure}[t!]
  \centering
  \includegraphics[width=1.0\columnwidth]{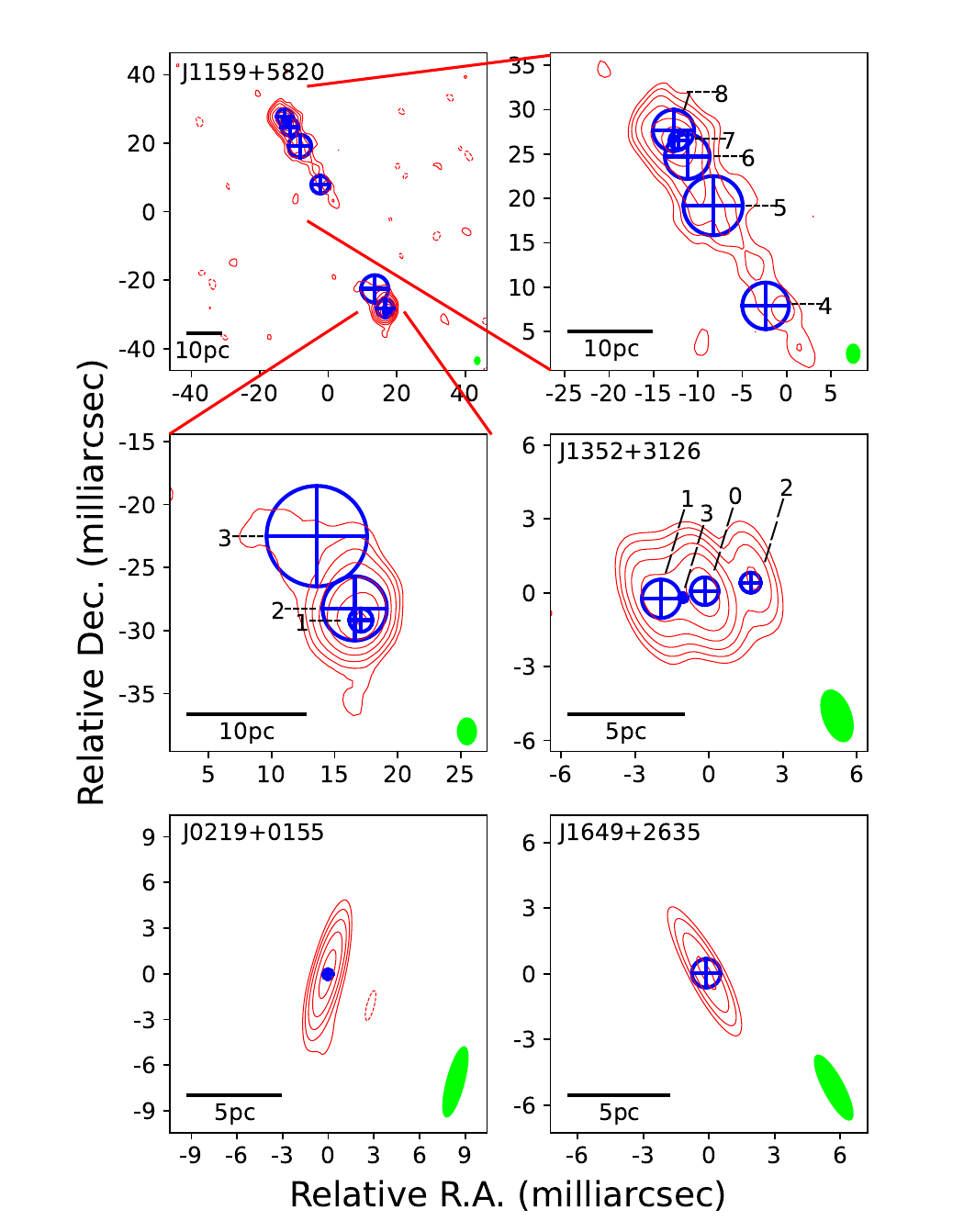}
  \caption{
  Distribution of the Gaussian components fitted to the VLBA data. 
  In all panels, contours show the total intensities and blue circles with crosses indicate the center locations FWHM sizes of the Gaussians with their corresponding IDs (see Table \ref{tab:BT}).
  Green ellipses in the bottom right are the corresponding VLBA beams.
  The source names are indicated on the upper left corners. 
  Top right and middle left panels are zoomed-in crops of the top left panels. 
  In all panels, the contours correspond to ($-$1, 1, 2, 4, 8, 16, 32, 64, 128, 256)$\times 3\sigma$ of the individual images.
  }
  \label{fig:gaussian_fitting_imgaes}
\end{figure}

\begin{table*}[t]
  \centering
  \caption{
  Parameters of the Gaussian model components fitted to the VLBI datasets. 
  }
  \begin{tabular}{ccccccc}
    \hline
    Target & ID & $F$ & $\Theta$ & $T_{\rm b}$ & $r$ & $PA$ \\
    (J2000)&    & (mJy) & (mas) & $(10^{9} \,K)$ & (mas) & (deg) \\
    \hline
    J0219+0155 & 0 & 22.2$\pm$18.6 & 0.47$\pm$0.03 & 1.6$\pm$1.4 & - & - \\
    \hline
    J1159+5820$^\dagger$ & 1 & 50.7$\pm$9.1 & 1.3$\pm$0.1 & 1.6$\pm$0.4 & - & - \\
               & 2 & 82.1$\pm$11.6 & 3.6$\pm$0.4 & 0.33$\pm$0.09 & 1.1$\pm$0.2 & 31.8$\pm$8.8 \\
               & 3 & 8.0$\pm$4.1 & 5.6$\pm$2.8 & 0.013$\pm$0.002 & 8.2$\pm$1.4 & 33.8$\pm$9.6 \\
               & 4 & 9.9$\pm$3.1 & 3.7$\pm$1.0 & 0.04$\pm$0.02 & 41.2$\pm$0.5 & 25.9$\pm$0.7 \\
               & 5 & 32.6$\pm$7.2 & 4.7$\pm$0.9 & 0.08$\pm$0.03 & 54.9$\pm$0.5 & 28.3$\pm$0.5 \\
               & 6 & 38.6$\pm$6.8 & 3.6$\pm$0.5 & 0.16$\pm$0.05 & 60.8$\pm$0.3 & 27.7$\pm$0.2 \\
               & 7 & 51.7$\pm$8.6 & 1.5$\pm$0.1 & 1.3$\pm$0.3 & 63.03$\pm$0.05 & 27.91$\pm$0.03 \\
               & 8 & 45.6$\pm$7.3 & 3.3$\pm$0.4 & 0.22$\pm$0.06 & 64.9$\pm$0.2 & 28.8$\pm$0.2 \\
    \hline
        J1352+3126 & 0 & 37.8$\pm$9.5 & 0.55$\pm$0.03 & 2.1$\pm$0.6 & - & - \\
               & 1 & 16.9$\pm$5.4 & 0.63$\pm$0.04 & 0.7$\pm$0.2 & 1.97$\pm$0.02 & 97.6$\pm$0.4 \\
               & 2 & 12.3$\pm$6.3 & 0.44$\pm$0.03 & 1.0$\pm$0.6 & 2.13$\pm$0.01 & -69.5$\pm$0.2\\
               & $3^{\dagger\dagger}$ & 5.6$\pm$8.4 & $<0.22$ & $>1.9$ & 0.93 & 85.9$\pm$0.2 \\
    \hline
    J1649+2635 & 0 & 16.0$\pm$5.4 & 0.93$\pm$0.09 & 0.3$\pm$0.1 & - & - \\
    \hline
  \end{tabular}
  \label{tab:BT}
  \tablecomments{ 
  From left, each column shows the source names in J2000, 
  the Gaussian model component IDs corresponding to the numbering in Fig.~\ref{fig:gaussian_fitting_imgaes} (0 indicating the VLBI core), 
  flux densities in mJy, 
  FWHM sizes in mas, 
  the observed brightness temperatures in unit of $10^9$\,K,
  the relative angular separations from the highest $T_{\rm b}$ component in mas, 
  and the position angles measured east of north in degrees.
  $\dagger$: J1159+5820 was observed at 4.8\,GHz while all other objects were observed at $8.4-8.6$\,GHz.
  $\dagger\dagger$: This component was included in the fitting process, but its detection is only marginal, given the faintness and ill-constrained size.
  }
\end{table*}

Figure~\ref{fig:all_images} shows composite optical and radio images for each source, spanning spatial scales from kpc to compact pc-scales.
For J1159+5820 and J1352+3126, SDSS and LoTSS DR2 (J1159+5820) and HST and LoTSS DR2 (J1352+3126) images are presented. 
Visual inspection of the HST images provides clearer views of the disk and dust-lane geometry of host galaxies in optical wavelengths, in particular for J1352+3126, which strengthens the case for a spiral host galaxy.
In radio wavelengths, deeper and higher-resolution views of radio structures on kpc-scales are also available.
We find that the kpc-scale radio jets in J0219+0155 and J1159+5820 show X-like shape, indicating potentially complex dynamics or evolution of these jets (e.g., \citealt{capetti02,joshi19}).
More quantitative details of the large-scale radio images are summarized in Table \ref{tab:VLBA_data_summary}.
On the pc-scales, all the four sources show at least compact core-like structure. 
Furthermore, two objects (J1159+5820 and J1352+3126) clearly exhibit extended structure, which we regard as jet features, although the identification of the VLBI core and furthermore the central BH is not straightforward with single-frequency only images, considering effects due to synchrotron opacity (e.g., \citealt{Lobanov98}). 
We note that all the targets are relatively nearby objects ($z < 0.1$) and the detected milliarcsecond-scale radio components could also be associated with star-formation processes. 
Nevertheless, multi-wavelength properties of the targets and also the brightness temperatures (see below) support the AGN origin of the compact radio emission.
Below we provide more specific details for each source.
As for J0219+0155 and J1649+2635, their natures were identified as radio-loud active galaxies based on the ratios of the far infrared to radio brightness \citep{condon11}. 
Likewise, VLBI morphology of the source (see Fig. \ref{fig:all_images}) and the compact symmetric object (CSO) classification of the object \citep{Tremblay16} support the AGN-origin of the radio structure.
Lastly, the pc-scale structure in J1352+3126 was also identified as being associated with AGN based on the extended jet-like morphology \citep{Beswick04,Giovannini05}.
We further discuss definition and possible identifications of the VLBI core in Sect.~\ref{sec:structure_identification} in Appendix.
We also note that the position angles of the extended jets in the two objects are significantly different on pc- and kpc-scales, by $\sim45^{\circ}$. 

In Fig.~\ref{fig:gaussian_fitting_imgaes}, we present the results of the Gaussian model fitting to the VLBA data, with locations of the fitted components overlaid on the VLBI images. 
Detailed parameters of each component are listed in Table~\ref{tab:BT}. 
Overall, the VLBI cores are relatively faint, with characteristic flux densities of a few tens of mJy (see Table \ref{tab:BT}).
More importantly, all the four sources exhibit $T_{\rm b}$ of order of $10^9$\,K. 
While these values far exceed $\sim10^{5}\,$K, roughly an upper threshold value for radio emission of starburst origin (e.g., \citealt{Condon91}), the observed $T_{\rm b}$ are still low for radio AGNs, compared to the typical range of $T_{\rm b} \sim 10^{11}-10^{12}$\,K observed from the cores of compact jets in typical quasars and blazars at 15\,GHz \citep{Homan21}.
Although $T_{\rm b}$ in jets of AGNs have slight frequency dependence ($T_{\rm b}\propto \nu^{-0.95}$; \citealt{roeder25}), the frequency difference between $2-8\,$GHz and $15\,$GHz is small compared to the substantial factor $\sim10^{2}-10^{3}$ difference. 
The values of $T_{\rm b}\sim10^{9}\,$K are also substantially lower than the typical brightness temperature at the kinetic and magnetic energy equipartition; $T_{\rm b,eq}\sim5\times10^{10}$\,K \citep{Readhead94}.
The low values of observed $T_{\rm b}$ imply that relativistic effects are not dominant in these sources, unlike strongly beamed AGN jets such as those in blazars. Also, the detection of the likely jet and counterjet structures in the VLBI images (see below and Sect. \ref{sec:structure_identification} in Appendix) further supports an intrinsic source geometry close to the plane of the sky, so that the low $T_{\rm b}$ is most naturally explained by intrinsically faint radio emission from a low-power relativistic outflow near the core.

We also show the result of our calculations of $\log(L_{\rm{R,1.4\,GHz}}/L_{\rm Edd})$ for the four objects in Table~\ref{tab:properties_of_targets}.
All the sources show $\log(L_{\rm{R,1.4\,GHz}}/L_{\rm Edd})\sim-(5-8)$; these are small numbers compared to the result of \cite{King17}, who show that logarithms of radio-Eddington ratios are typically $\gtrsim -4$ for powerful blazars and quasars but $\lesssim-7$ for Seyfert galaxies without large kpc/Mpc-scale jets. 
For J1159+5820, we note that an independent estimate of the bolometric Eddington ratio based on the H$\alpha$ luminosity is available in the literature, yielding $L_{\rm bol}/L_{\rm Edd} \sim 10^{-3}$ \citep{Balasubramaniam20}.
This value is significantly larger than typical values found in traditional low-luminosity AGNs with jet features 
(e.g., \citealt{Yuan14}), and may disfavor the dominance of a pure radiatively inefficient accretion flow and powerful jet in J1159+5820 and likely also in other spiral DRAGN cores.
These comparisons again support the currently weak powers of jets in spiral DRAGNs on pc-scales, compared to past state which was responsible for producing the large-scale lobe structures.

\begin{figure}[t!] 
  \centering
  \includegraphics[width=1.0\columnwidth]{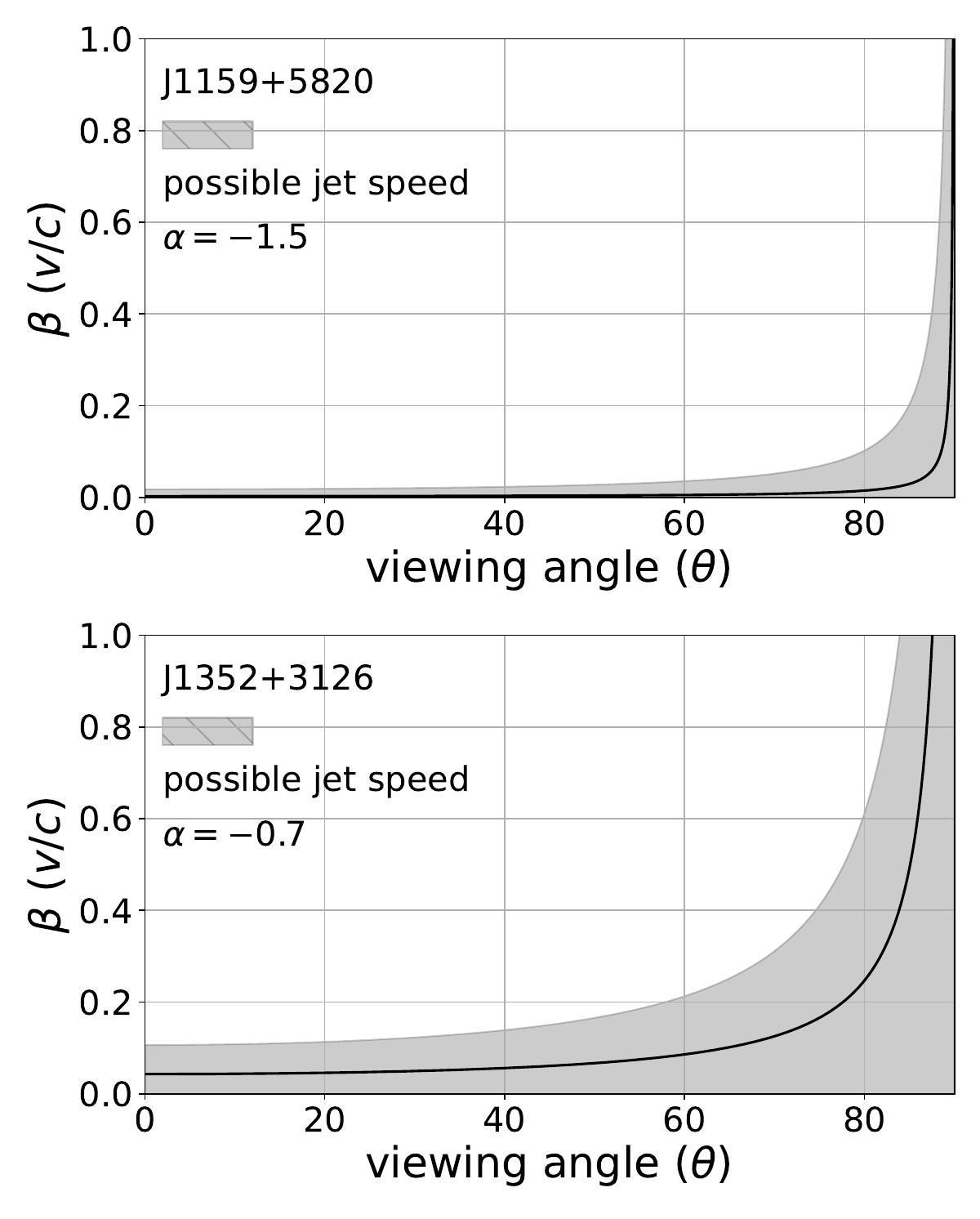}
  \caption{
  Possible jet speeds of (upper) J1159+5820 and (lower) J1352+3126, derived based on the scenarios outlined in the main text and Sect.~\ref{sec:structure_identification}. 
  In both panels, the black solid line and the gray regions represent best-fit and allowed ranges of the jet speeds including uncertainties in $R_{\rm jet}$ ($1\sigma$ level), respectively}. 
  \label{Fig:jet_intrinsic_speed}
\end{figure}

We also make attempts to constrain the intrinsic jet speeds for J1159+5820 and J1352+3126 which show resolved jet-like features. 
To do so, we need to consider multiple possible identifications of the VLBI core and jet components for the calculation of the jet-to-counterjet brightness ratio.
The detailed procedure for identifying jet and counterjet components is described in Sect.~\ref{sec:structure_identification} of Appendix.
We estimated the intrinsic jet speed, $\beta$, using the jet–to–counterjet brightness ratio,
$R_{\rm jet} = F_{\rm J}/F_{\rm CJ}$,
where $F_{\rm J}$ and $F_{\rm CJ}$ are the model-fitted flux densities.
For $F_{\rm J}$ ($F_{\rm CJ}$) of J1159+5820, we used the summed flux densities for components 6, 7, and 8 (components 1 and 2), respectively.
For J1352+3126, we simply made use of the components 1 and 2.
Relativistic beaming  boosts the observed flux density of a moving blob as
$F_{\rm obs} = \delta^{3-\alpha} F$,
where $F_{\rm obs}$ and $F$ are the observed and intrinsic flux densities and $\alpha$ is the spectral index $(F_\nu \propto \nu^{+\alpha})$.
The Lorentz and Doppler factors are then
$\gamma = 1/\sqrt{1-\beta^2}$ and $\delta = 1/[\gamma(1-\beta\cos\theta)]$, 
where $\theta$ is the viewing angle.
The resulting intrinsic jet speed can be expressed in observational terms as 
$\beta = (R_{\rm jet}^{1/(3-\alpha)}-1)/[(R_{\rm jet}^{1/(3-\alpha)}+1)\cos\theta]$ \citep{Kim18}.
We adopted a spectral index of $\alpha \approx -1.5$ for J1159+5820 \citep{Tremblay16} and $\alpha \approx -0.7$ for J1352+3126.
The estimated jet-to-counterjet brightness ratios are $R_{\rm jet}=1.02\pm0.15$ for J1159+5820 and $R_{\rm jet}=1.37\pm0.83$ for J1352+3126.
Based on these numbers, we show the results in Fig.~\ref{Fig:jet_intrinsic_speed} for the resulting ranges of intrinsic jet speeds, accounting for uncertainties in $R_{\rm jet}$.
For J1159+5820, we find that $\beta\gtrsim0.10$ is allowed only when the jet viewing angle is substantially large with $\theta>80^{\circ}$; otherwise the object appears to be at most only mildly relativistic, consistent with the expectation based on the low $T_{\rm b}$ value.
As for J1352+3126, somewhat faster and more relativistic jet speeds are allowed, for instance, $\beta\gtrsim0.6$ ($\Gamma\gtrsim1.25$) for the same $\theta>80^{\circ}$. 
In both cases, we find $\beta\lesssim0.6$ for a moderate range of $\theta\lesssim80^{\circ}$.
Although the lack of independent constraints on $\theta$ makes it difficult to draw a firm conclusion (see also the absence of correlation between the position angles of the jets and galaxy disks in spiral DRAGNs; \citealt{Wu22}),   
the combination of the above $\Gamma$ values and the plausible range of $\theta$ still points to a mildly, or at most only moderately, relativistic outflow, consistent with the observed $T_{\rm b}$ and low $\log(L_{\rm R,1.4\,GHz}/L_{\rm Edd})$.

\begin{figure}[t] 
  \centering
  \includegraphics[width=1.0\columnwidth]{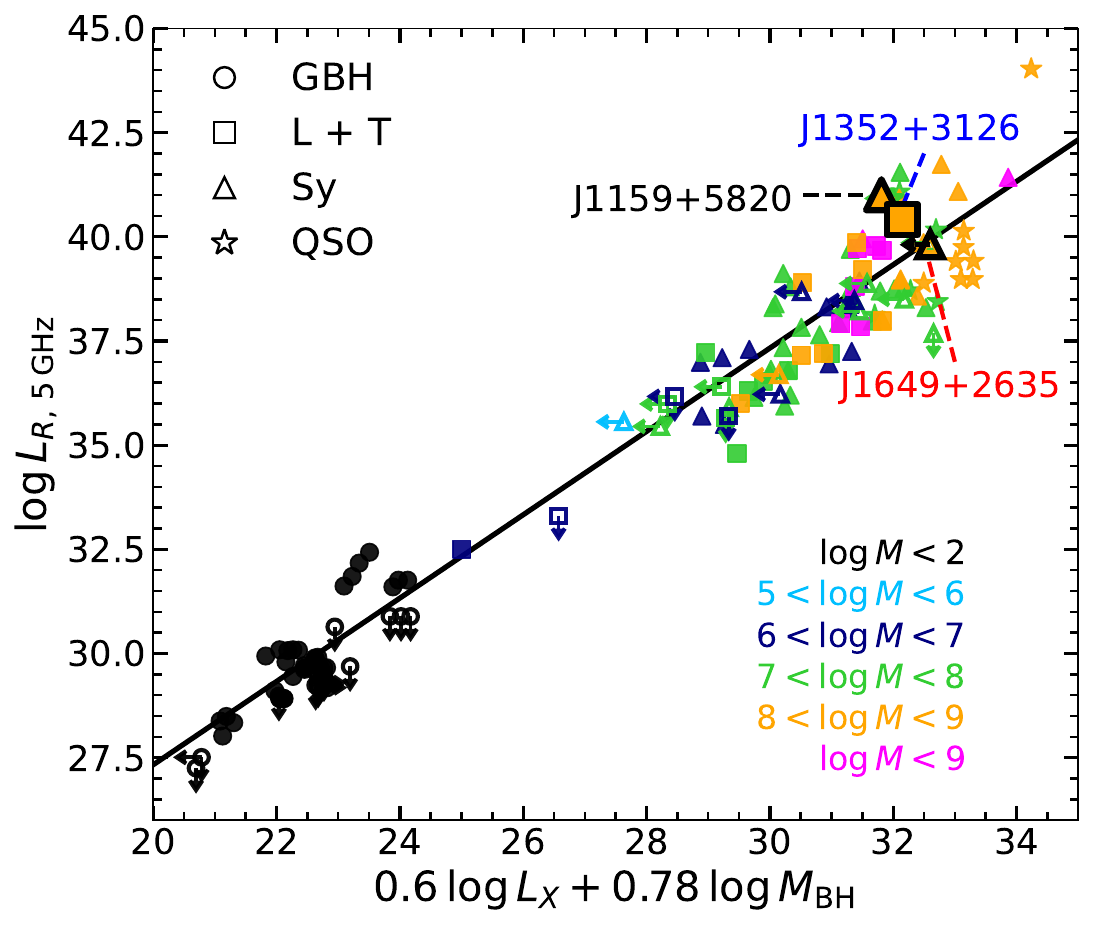}
  \caption{
  The fundamental plane of active black holes \citep{Merloni03}, including data from \cite{Merloni03} and those of three of four targets in our study. The latter is highlighted by error bars and colored dashed lines with corresponding source names.  
  Upper limits are marked with arrows and shown in empty symbols.
  The color denotes ranges of $M_{\rm BH}$.
  Different symbols show types of the BHs---`GBH': Galactic black holes, `Sy': Seyfert galaxies, `L': Low-ionization nuclear emission-line regions, `T': Transition objects (Liner / H\,II),  `QSO': Quasi-stellar objects.  
  }
  \label{Fig:fundemantal_plane}
\end{figure}

Lastly, in Fig.~\ref{Fig:fundemantal_plane} we show the locations of three spiral DRAGN objects in the fundamental plane of active black holes (Sect.~\ref{subsec:other_mwl_data}), using the values of $M_{\rm BH}$, $L_{\rm 2-10\ keV}$, and $L_{\rm R,5\,GHz}$.
Despite their unusual large-scale radio morphology and spiral hosts, all three objects lie consistently on the established relation  $\log L_{\rm R,5GHz}=0.6\log L_{\rm X} + 0.78\log M_{\rm BH}$ \citep{Merloni03}, similar to other BH systems ranging from stellar mass objects to AGNs. 
This may imply that global accretion-jet coupling in spiral DRAGNs follows the same underlying physical scaling seen in broader AGN population, suggesting that spiral DRAGNs do not have fundamentally different central engine or jet launching mechanism (see also \citealt{Lee25} for more details about the dynamics of jets in spiral DRAGNs).

\section{Discussions}\label{sec:discussion}

As in the case of 0313--192 \citep{Mao18}, our detection of compact, VLBI-scale radio structures in these spiral DRAGNs strengthens the association between the nuclear radio emission and their optical spiral hosts. This firmly supports the serendipitous identification of all four targets. Below, we discuss the physical implications of these findings.

Most importantly, the pc-scale outflow structures detected in J1159+5820 and J1352+3126 show that jet production in these systems is presently active, or has been reactivated in the recent past. 
Such ongoing nuclear activity is difficult to reconcile with scenarios in which spiral DRAGNs are merely relics of a single, short-lived, merger-triggered outburst \citep{Ledlow98,Singh15,Mao18}. 
Instead, the coexistence of active pc-scale cores with large, FR~II-like outer lobes, together with the known double-double (J1352+3126, \citealt{saikia09}) and X-shaped (J0219+0155, \citealt{Yang19}; J1159+5820, \citealt{misra23}) kpc-scale morphologies of the sources in our sample, strongly suggests multiple episodes of jet launching. 
The VLBI detections therefore indicate that the current nuclear activity represents a new cycle following the episode that inflated the outer lobes.
Recent high-resolution hydrodynamical simulations also reinforce this broader picture. 
In particular, simulations of gas-rich major mergers that include spin-driven AGN jets show that while a major merger can trigger a powerful jet episode, the interaction naturally generates large-scale multiphase outflows, the cooling and fallback of gas, and renewed fueling of the SMBH, thereby producing recurrent and sometimes misaligned jet episodes \citep{talbot24}. 
These results demonstrate that even in systems where a gas-rich major merger occurs, the associated jet activity is not restricted to a single outburst. Instead, the long-term duty cycle is governed by the availability of returning or secularly supplied gas and by the episodic build-up of magnetic flux near the SMBH.

In this context, models in which the long-term evolution of the jet is regulated by internal properties of the host galaxy, such as the connection between the stellar bulge and the black hole or the presence of sustained low-level gas inflow, provide a natural explanation for recurrent nuclear activity. 
Such mechanisms reduce the reliance on a single major merger as the unique trigger and are consistent with intermittent or stochastic accretion that can power multiple jet episodes over Gyr timescales \citep{Wu22}.
This also better agrees with what is commonly inferred from double–double radio galaxies, where the large-scale radio lobes are interpreted as relics of previous jet activity--for example as seen in J1352+3126 \citep{saikia09}. 

In contrast, the interpretation about the two other sources J0219+0155 and J1649+2635, showing compact and barely resolved nuclear radio emission, can be diverse. 
They may be powered in different manner, such as intrinsically weaker and compact jets or radiatively inefficient accretion flows 
\citep[e.g.,][]{Mahadevan98,Yuan03,Narayan05,Yuan14}. This is similar to the origin of radio emission in low-luminosity AGNs (LLAGNs; \citealt{Ho08, Gutierrez21}).
If these objects are indeed in a currently low-activity phase, their kpc-scale jet structure could indeed reflect an earlier single jet outburst. 
We also note the lack of any additional larger-scale diffuse radio emission beyond the kpc-scale radio lobes in J1649+2635 \citep{mao15}, which is not against our speculation.
On the other hand, we cannot exclude the possibility that faint pc-scale jets in the two objects are simply missing due to the current sensitivity or resolution limits of the VLBA.
Therefore, we may conservatively conclude at this point that the origin of jets in spiral DRAGNs is unlikely to arise from a single, uniform mechanism. 
We also note examples of kpc-scale multiple jet structures in other spiral DRAGNs supporting this interpretation; see J1409-0302 \citep{hota11} and J2345-0449 \citep{bagchi14}.
Accordingly, we may suggest that spiral DRAGNs in general may span a range of evolutionary states, from presently active jets to sources in quiescent or weakly accreting phases. This implies a diversity of pathways that can lead to large-scale radio emission in late-type host galaxies.

Regarding the jet dynamics, the estimated low intrinsic jet speeds ($\beta \lesssim 0.8$), 
small radio-Eddington ratios of $\log(L_{\rm R, 1.4\,GHz}/L_{\rm Edd}) \simeq -(5-8)$, and low brightness temperatures of the VLBI scale cores ($T_{\rm b} \approx 10^9$~K) indicate that the pc-scale outflows in spiral DRAGNs are only mildly relativistic and far from the highly accelerated and boosted jets seen, for example, in blazars. 
We note that traditional hot accretion flow-based general relativistic magnetohydrodynamics (GRMHD) models of jet formation generally predict only mildly relativistic outflows near the jet base, which are then magnetically accelerated up to large kpc-scales \citep{Blandford19}. Similarly, recent GRMHD simulations of jets launched from thin or moderately thick disks \citep{Liska20} also exhibit comparatively slow outflows at small $\lesssim100r_{\rm g}$ radii from the central SMBHs where $r_{\rm g}=GM_{\rm BH}/c^{2}$ is the gravitational radius. 
Interestingly, such jets can then be substantially accelerated up to terminal $\Gamma\sim10$, 
after a substantial reservoir of large-scale vertical magnetic flux has accumulated near the disk.
Because this flux is  built gradually through an $\alpha-\Omega$ dynamo \citep{Parker55, Moffatt78}, the jet may be initially slow and develop into a fast relativistic outflow only at later stages. 
In this regard, the observed VLBI cores in spiral DRAGNs may trace an early or intermittently magnetized phase of the jet, while the kpc-scale lobes preserve the signatures of earlier, more powerful outbursts.

We also note that, whereas previous studies have attributed the rarity of spiral DRAGNs to environmental effects on large galactic scales (e.g., \citealt{Singh15, Wu22}), our study focused on the innermost regions suggests an additional possibility that the rarity and apparent intermittency of spiral DRAGNs may instead reflect the difficulty of building and maintaining the vertical magnetic fields required in the immediate vicinity of the black hole.
As shown by \citet{Liska20}, reaching and maintaining a magnetically arrested state demands both prolonged accretion and the retention of large-scale poloidal flux near the inner disk. 
Such conditions are difficult to be achieved in spiral galaxies, where gas inflow is typically clumpy, time-variable, and largely driven by secular processes rather than major mergers \citep{Wada02,Hopkins10}.
In these environments, the vertical magnetic field would originate from turbulent, predominantly toroidal fields generated within the disk, and it would be built up into poloidal fields only episodically. 
For instance, short-lived episodes of enhanced inflow could temporarily strengthen the vertical field and power a more luminous jet, while subsequent dissipation of magnetic flux or reconnection events would return the system to a low-activity state \citep{Cielo18}.
This cycle naturally leads to long quiescent intervals between short periods of jet activities. 
For example, the spectral aging analysis of J1159+5820 and J1352+3126 yields lobe ages of several tens of Myr \citep{Joshi11,Machalski16,misra23}, indicating long quiescent intervals between individual jet-launching episodes. 
This may help explain both the small number of known spiral DRAGNs and diverse, often irregular morphologies
on large kpc-scales, such as those observed in J1352+3126 (double-double radio galaxy; \citealt{saikia09}) and J0219+0155 and J1159+5820 (X-shaped structures; \citealt{Yang19,misra23}). 
They may be direct evidence for short and recurrent phases of renewed jet activity following the aforementioned inactive periods.

\section{Conclusion}

In summary, we suggest that spiral DRAGNs, despite their rarity and serendipity, could be not a fundamentally separate class of SMBHs, but instead they may represent intermittently fueled regime of the same objects with fundamentally the same accretion-jet coupling. 
Their low-$T_{\rm b}$ VLBI cores and indications of slow jet speed support the idea that the inner spiral DRAGN jets are slow, magnetically under-developed phase of relativistic jet, whereas their largely extended radio lobes correspond to their earlier, more strongly magnetized outflows.
This can also help explain why faint pc-scale cores, weak jets, and powerful kpc-scale lobes can co-exist in these late-type host galaxies.
Future high-sensitivity VLBI surveys and polarimetric imaging of more spiral DRAGN candidates, for instance using the Square Kilometre Array, will enable more complete sample studies of these enigmatic objects, to test whether changes in the magnetic flux of the inner region drive recurrent jet formation in host galaxies of various types.

\begin{acknowledgements}
We thank the anonymous reviewer for constructive comments and suggestions, which helped improve the manuscript.
The authors thank Leonid Petrov and Gregory Taylor for permission to use some of their VLBA data for this study.
This work is supported by the National Research Foundation of Korea (NRF) grant funded by the Korean government (Ministry of Science and ICT; grant no. 2022R1C1C1005255, RS-2022-NR071771, 
RS-2024-00466005)
and by the Korea Astronomy and Space Science Institute under the R\&D program (Project No. 2025-9-844-00) supervised by the Korea AeroSpace Administration.
We acknowledge the use of the Astrogeo VLBI FITS image database hosted at \url{http://astrogeo.smce.nasa.gov/vlbi_images/} and maintained by Leonid Petrov.
This research used the facilities of the Canadian Astronomy Data Centre operated by the National Research Council of Canada with the support of the Canadian Space Agency.
This research is based on observations made with the NASA/ESA Hubble Space Telescope obtained from the Space Telescope Science Institute, which is operated by the Association of Universities for Research in Astronomy, Inc., under NASA contract NAS 5–26555. These observations are associated with programs hst\_15445\_3y\_acs\_wfc\_f475w\_jds43y and u27l5a02t.
This research has made use of the CIRADA cutout service at URL cutouts.cirada.ca, operated by the Canadian Initiative for Radio Astronomy Data Analysis (CIRADA). CIRADA is funded by a grant from the Canada Foundation for Innovation 2017 Innovation Fund (Project 35999), as well as by the Provinces of Ontario, British Columbia, Alberta, Manitoba and Quebec, in collaboration with the National Research Council of Canada, the US National Radio Astronomy Observatory and Australia’s Commonwealth Scientific and Industrial Research Organisation.
Funding for the Sloan Digital Sky Survey V has been provided by the Alfred P. Sloan Foundation, the Heising-Simons Foundation, the National Science Foundation, and the Participating Institutions. SDSS acknowledges support and resources from the Center for High-Performance Computing at the University of Utah. SDSS telescopes are located at Apache Point Observatory, funded by the Astrophysical Research Consortium and operated by New Mexico State University, and at Las Campanas Observatory, operated by the Carnegie Institution for Science. The SDSS web site is \url{www.sdss.org}.
SDSS is managed by the Astrophysical Research Consortium for the Participating Institutions of the SDSS Collaboration, including the Carnegie Institution for Science, Chilean National Time Allocation Committee (CNTAC) ratified researchers, Caltech, the Gotham Participation Group, Harvard University, Heidelberg University, The Flatiron Institute, The Johns Hopkins University, L'Ecole polytechnique f\'{e}d\'{e}rale de Lausanne (EPFL), Leibniz-Institut f\"{u}r Astrophysik Potsdam (AIP), Max-Planck-Institut f\"{u}r Astronomie (MPIA Heidelberg), Max-Planck-Institut f\"{u}r Extraterrestrische Physik (MPE), Nanjing University, National Astronomical Observatories of China (NAOC), New Mexico State University, The Ohio State University, Pennsylvania State University, Smithsonian Astrophysical Observatory, Space Telescope Science Institute (STScI), the Stellar Astrophysics Participation Group, Universidad Nacional Aut\'{o}noma de M\'{e}xico, University of Arizona, University of Colorado Boulder, University of Illinois at Urbana-Champaign, University of Toronto, University of Utah, University of Virginia, Yale University, and Yunnan University.
\end{acknowledgements}

\appendix
\section{Detailed properties of each object}\label{sec:details_all_targets}

Below we describe detailed properties and remarks for each of the four objects discussed in the main text.

\noindent
--J0219+0155: 
This source is a nearby galaxy with a spectroscopic redshift of z = 0.041, and optical images reveal a pronounced bulge together with distinct dust structures \citep{Wu22}. 
\cite{Yang19} classified the radio source as a winged or X-shaped system, based on the FIRST image shows low surface brightness lateral extensions emerging symmetrically from the central region on both sides, forming secondary structures that are characteristic of FR\,I type X-shaped radio galaxies such as NGC 326. 
When its optical magnitude and radio luminosity are compared with the empirical FR I/FR II boundary proposed \citep{Ledlow96}, the source lies near the FR\,I regime, although the extended kpc-scale emission shows FR\,II like outer features in the FIRST image \citep{Wu22}.
Structural decomposition yields bulge to total luminosities $B/T \approx 0.31$ and a single-component S{\'e}rsic index of $n = 2.01$ \citep{Wu22}, values that are consistent with a disk-dominated host containing a modest bulge. 

\noindent
--J1159+5820: 
Deep optical imaging reveals a prominent tidal tail and shell-like debris, indicating that this source is a post-merger system \citep{misra23}. 
\cite{Koziel12} and \cite{misra23} further described its radio morphology as X-shaped based on VLA and GMRT images, noting the presence of low surface brightness wings that extend away from the primary jet axis. 
\cite{Singh15} interpreted the kpc-scale radio structure using NVSS and FIRST images as evidence for multiple cycles of jet activity, based on the combination of an inner FR\,II like pair of lobes embedded within larger outer lobes. 
Therefore, the system was classified as an episodic AGN. 
\cite{Singh15} also reported a high $B/T \approx 0.75$ and an unusually large S{\'e}rsic index of $n \simeq 7.7$. 
However, such a high value should not be interpreted as evidence for an unusually concentrated, early-type morphology because the tidal tails and shell features identified by \cite{misra23} indicate a strongly disturbed surface brightness distribution that can artificially inflate the S{\'e}rsic index of the systems undergoing or following a merger.
\cite{Tremblay16} found that the central pc-scale emission exhibits a CSO morphology, and spectral aging studies yield a characteristic age of $\sim40-60$\,Myr for the outer lobes \citep{misra23}.

\noindent
--J1352+3126: 
This galaxy has been classified as both a spiral and an irregular galaxy (\citealt{Singh15} and references therein). 
Structural measurements give a high $B/T \approx 0.61$ and a comparatively high S{\'e}rsic index of $n = 5.83 \pm 0.27$ \citep{Singh15}.
However, we also note that the high-resolution HST image reveals the complexity in its central structures, suggesting that the ongoing interaction may be disturbing the stellar light distribution and, in turn, inflating the measured S{\'e}rsic index.
The host is rich in cold molecular gas \citep{Evans99}, with CO observations revealing a substantial reservoir of $M(\rm {H_2}) = 2.2 \times 10^{10}\,M_\odot$ \citep{Labiano14}.
VLA and GMRT imaging reveal an older, edge-brightened FR\,II like outer double with strongly asymmetric lobe brightness \citep{Joshi11}, as well as a younger, smaller inner double aligned along the same axis and accompanied by diffuse off-axis extensions \citep{Machalski16}.
Therefore, this source has been classified as a double–double radio galaxy  \citep{saikia09}. 
The spectral-age contrast between the outer ($\leq17-23\,\rm Myr$) and inner ($\leq0.1\,\rm Myr$) doubles provides additional support for interpreting the system as having undergone recurrent episodes of jet activity \citep{Joshi11, Machalski16}.
On parsec scales, earlier VLBI observations identified a flat spectrum core together with sub-kpc jet components \citep{Beswick04, Giovannini05}.

\noindent
--J1649+2635: 
This system is a grand-design spiral galaxy with prominent arms, an extended stellar halo, and a faint central bar \citep{mao15}. 
\cite{mao15} also noted the presence of a warped stellar disk, which they interpreted as evidence that the galaxy likely experienced a minor merger in the past, although the global spiral pattern remains largely preserved. 
Structural measurements give a  $B/T \approx 0.63$ and a S{\'e}rsic index of $n = 2.81 \pm 0.03$ \citep{Singh15}. 
Optical spectroscopy from the SDSS shows a red continuum and a strong 4000\,\AA\ break, indicating predominantly quiescent star formation \citep{Masters10}. 
VLA and MERLIN radio imaging further reveal edge-brightened outer lobes characteristic of an FR\,II radio morphology (see \citealt{mao15} and references therein).

\section{Identification of VLBI-scale Structure}\label{sec:structure_identification}

Here, we describe the procedure and details of the identification of VLBI-scale core, jet, and counterjet components in J1159+5820 and J1352+3126.
For J1159+5820, we adopt a jet–counter-jet (J–CJ) scenario; that is, the two separated components represent the approaching (brighter) and receding (fainter) jets, while the VLBI core (or apparent footprint of the jet; \citealt{Lobanov98}) is not visible in the image due to the large opacity surrounding the central engine. 
This is consistent with the previous classification of the source as a CSO based on the comparable brightness of the two brightest features in the VLBI image \citep{Tremblay16}.
Based on the apparent brightness, we may identify the NE component as being associated with the approaching jet and the SW component as being associated with the receding jet.
We note that, in principle, the brightness ratio of the NE and SW features could be affected not only by the beaming but also by other factors. For instance, the synchrotron or free-free absorption could affect the observed brightness ratio, but for J1159+5820, both jet ends display steep spectra at 5-8\,GHz, and thus are expected to be not strongly affected by the absorption effect. 
Another possible factor contributing to the observed brightness asymmetry is jet–environment interaction, which can significantly affect the jet-to-counterjet flux ratio.
Such effects can also be further assessed by comparing the arm-length ratio between the brighter and fainter lobes (with values $>1$ generally indicating a weaker environmental influence). 
Although the absence of a clearly identifiable VLBI core in this object \citep{Tremblay16} makes it difficult to measure this ratio reliably, the steep spectra at both ends again suggest weak jet-environment interaction. 
All in all, we consider it reasonable to associate the northern lobe with emission from the approaching jet, as it exhibits a longer visible tail of the jet extending southward, in contrast to the southern lobe, which shows only a faint or barely visible extension of the emission toward the north.
For J1352+3126, the high $T_{\rm b}$ feature between the two outer blobs of lower $T_{\rm b}$ supports the component being a VLBI core \citep[see, e.g.,][for general profiles of $T_{\rm b}$ as function of core distance for jets in AGNs]{burd22}.  
Thus, we can reasonably assume that the central brightest component corresponds to the core, with the eastern (western) brighter (fainter) feature corresponding to the jet (counter-jet) component.
The length ratio between the components 0 to 1 and 0 to 2 in J1352+3126 is $r_{0-1}/r_{0-2}\sim 0.93\pm0.01$ (Table~\ref{tab:BT}), which is not unreasonable for mild relativistic beaming with non-negligible environmental effects.

\bibliographystyle{aasjournalv7}
\bibliography{reference}{}

\end{document}